\documentclass[11pt, letterpaper]{JHEP3}
\usepackage{amssymb, amsmath, amsopn, amsthm}
\usepackage{epsfig}
\usepackage{psfrag}
\usepackage{subfigure}
\usepackage{cite}
\usepackage{epstopdf}
\usepackage{graphicx}
\usepackage{dcolumn}
\usepackage{bm, verbatim}

\newcommand{\beqa}{\begin{eqnarray}}
\newcommand{\eeqa}{\end{eqnarray}}
\newcommand{\be}{\begin{equation}}
\newcommand{\ee}{\end{equation}}
\newcommand{\bea}{\begin{eqnarray}}
\newcommand{\eea}{\end{eqnarray}}
\newcommand{\bA}{\begin{array}}
\newcommand{\eA}{\end{array}}
\newcommand{\bc}{\begin{center}}
\newcommand{\ec}{\end{center}}

\renewcommand{\thepage}{\arabic{page}} 

\def\coeff#1#2{{\textstyle {\frac {#1}{#2}}}}
\def\half{\coeff 12}

\def\N{{\cal N}}
\def\T{{\bf T}}

\def\S_1{{\widetilde {S_1}}}

\def\R{{\bf R}}
\def\S{{\bf S}}

\def\tr{{\rm tr}}

\def\x{\mathbf x}

\def\Z{{\mathbb Z}}

\def\Dslash{{\rlap{\raise 1pt \hbox{$\>/$}}D}}

\def\be{\begin{equation}}
\def\ee{\end{equation}}
\def\bea{\begin{eqnarray}}
\def\eea{\end{eqnarray}}

\title{  Deconfinement in Yang-Mills theory through \\ toroidal compactification with deformation}

\author{~{\normalsize \bfseries \sffamily Du\v san Simi\' c${}^{1,2}$ and Mithat \"Unsal${}^{1,2}$}

~\\${}^1$Department of Physics, Stanford University\\
Stanford, CA 94305 USA \\ \vspace{0.3cm}

${}^2$ Theory Group, SLAC National Accelerator Laboratory\\
Menlo Park, CA 94025 USA \\ \vspace{0.3cm}

\email{simic@stanford.edu, unsal@slac.stanford.edu}\\
}

\newcommand{\bbea}{\begin{equation} \begin{aligned}} \newcommand{\eeea}{\end{aligned} \end{equation}}

\abstract{ We introduce field theory techniques through which the deconfinement transition of  four-dimensional Yang-Mills theory can be moved to a semi-classical domain where it becomes calculable using two-dimensional field theory. We achieve this through a double-trace deformation of toroidally compactified Yang-Mills theory on $\R^2 \times \S^1_L \times  \S^1_\beta$. At large $N$, fixed-$L$, and arbitrary $\beta$, the thermodynamics of the deformed  theory  is equivalent to that of ordinary Yang-Mills theory at leading order in the large $N$ expansion. At fixed-$N$, small $L$ and a range of $\beta$, the deformed theory maps to a two-dimensional theory with electric and magnetic (order and disorder) perturbations, analogs of which appear in planar spin-systems  and statistical physics. 
 We show that in this regime the deconfinement transition is driven by the competition between electric and magnetic perturbations in this two-dimensional theory. This appears to support the scenario proposed by Liao and Shuryak [1] regarding the magnetic component of the quark-gluon plasma at RHIC. 
 }

\preprint{SLAC-PUB-14287}

\def\be{\begin{equation}}
\def\ee{\end{equation}}
\def\bea{\begin{eqnarray}}
\def\eea{\end{eqnarray}}

\begin{document}

\section{Introduction}

Through numerical simulations on the lattice  \cite{Aoki:2006we,Cheng:2006qk}   and the experimental program at the Relativistic Heavy  Ion Collider (RHIC) \cite{Arsene:2004fa, Adcox:2004mh, Back:2004je, Adams:2005dq}, we know that QCD has a high temperature deconfined quark-gluon plasma phase at temperatures above $T_c=\beta_c^{-1} \approx 170$ MeV, where $T_c$ is parametrically of the order of the strong scale of the theory. Through lattice simulations, it is also known that the pure gauge sector of QCD, Yang-Mills theory, has a low temperature confined phase and a high temperature deconfined phase \cite{Lucini:2005vg}. While symmetry and universality arguments are useful \cite{Svetitsky:1982gs}, to date, there is no direct  continuum field theory technique to address most aspects of this transition due to its non-perturbative nature.\footnote{ The deconfinement transition can also be studied by using strong coupling lattice models  \cite{Susskind:1979up}, however, it is not known how to extend this to phases continuously connected to the continuum. The existence of a deconfined phase (in continuum)  can be established in perturbation theory \cite{Gross:1980br}. } 

Our goal in this paper is to make progress in understanding the microscopic mechanism driving the deconfinement transition of QCD and related theories, hopefully providing new insights into the structure of the quark-gluon plasma in the temperature region around $T_c$. As reviewed in Ref.\cite{Muller:2006ee}, this is one of the important problems concerning the physics of the nuclear collisions at RHIC. 
  
Recently, two  new methods have been introduced for studying aspects of the deconfinement transition in a variety of gauge theories. The gauge/gravity correspondence, as realized by string theory, is a powerful tool for studying certain strongly coupled gauge theories \cite{Witten:1997ep, Witten:1998zw, Aharony:2006da}. The theories for which a semi-classical limit of string theory is useful usually differ from QCD in some way such as the existence of a non-decoupled KK-tower of states or an absence of asymptotic freedom. Regardless, this approach has the remarkable virtue of allowing one to do detailed calculations in a host of strongly coupled systems, many of which are plausibly in the same universality class as QCD or a QCD-like theory.  For some recent applications to finite temperature properties, see \cite{recent_AdS,Buchel:2009nx,Bigazzi:2009bk}. A second approach for studying deconfinement was developed in \cite{Aharony:2003sx, Aharony:2005bq, Unsal:2007fb, HoyosBadajoz:2007ds}, where one considers the large $N$ limit of four dimensional $SU(N)$ gauge theories compactified on $S^3 \times S^1$. For a small $S^3$, the theory reduces to a matrix model, and there is a calculable deconfinement transition. In this second approach the large $N$ limit is important for 
achieving the thermodynamic limit. Motivated by these two inspiring examples we pose the following questions:
\begin{quote}
 Can  we find a calculable deconfinement transition in an asymptotically free and  confining  gauge theory by using field theory techniques? Is this even possible as the transition itself is non-perturbative? Can we give a simple physical picture of the mechanism behind the deconfinement transition?
\end{quote}
The small $S^3 \times  S^1_\beta$ example provides an existence proof that finding calculable examples of deconfinement transitions in asymptotically free gauge theories is possible, at least at $N=\infty$ \cite{Aharony:2005bq, Unsal:2007fb, HoyosBadajoz:2007ds}.

In this work we introduce field theory techniques through which the deconfinement transition of  four-dimensional Yang-Mills theory can be moved to a semi-classical domain where it becomes calculable using two-dimensional field theory. We achieve this by studying a double-trace deformation of Yang-Mills theory on $\R^2 \times \S^1_L \times \S^1_\beta$, which we refer to 
as  the  ``$\Omega_L$-deformed Yang-Mills" theory or simply deformed Yang-Mills.  
 Our deformation is similar to the one studied in the context of large $N$ volume independence \cite{Unsal:2008ch}.  
 
\begin{figure}[t]
\begin{center}
\includegraphics[angle=-90, width=0.990\textwidth]{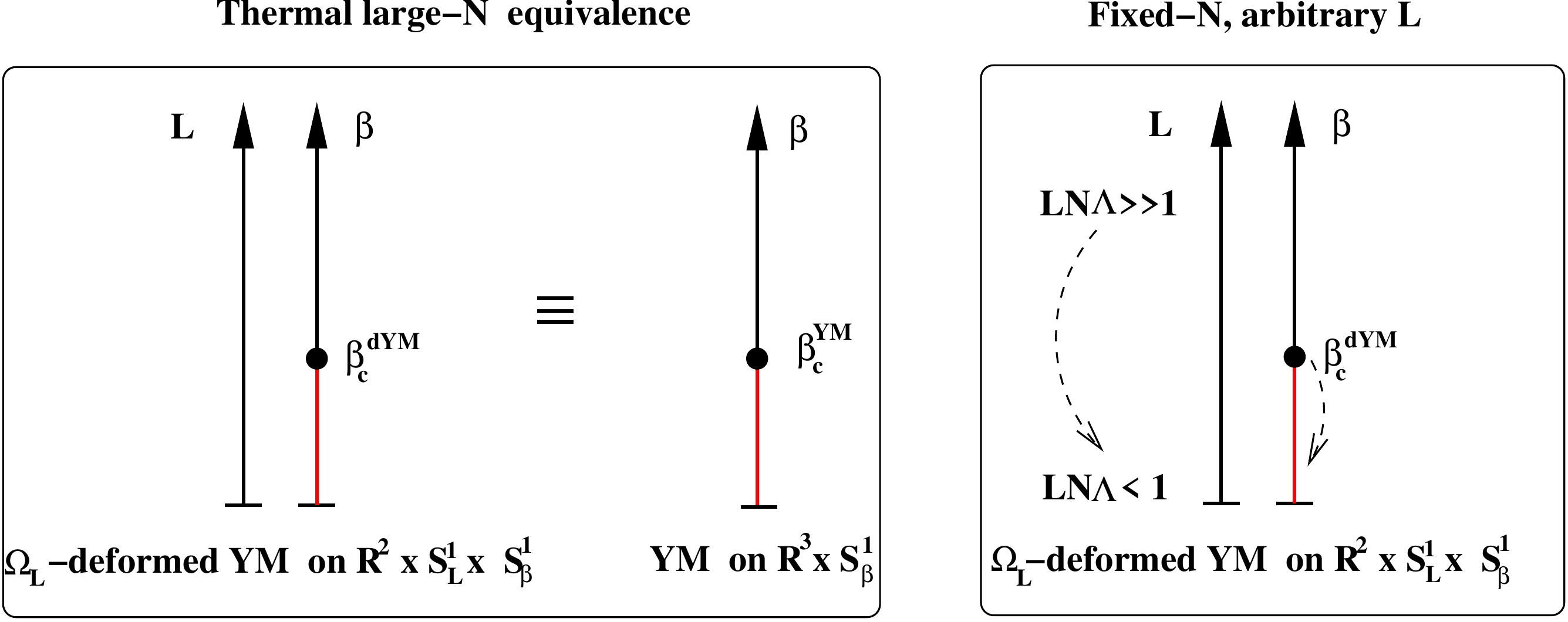}
\caption{ (Left) At $N=\infty$, the thermodynamics of the deformed and undeformed Yang-Mills theories are equivalent.  (Right) Unlike pure Yang-Mills, the deformed theory has, at finite $N$, a semi-classical domain (defined by $LN\Lambda \lesssim 1$) where the confinement-deconfinement transition is analytically calculable.   
}
  \label {fig:L-deform}
\end{center}
\end{figure}

The $\Omega_L$-deformation has the effect that at large $N$, fixed-$L$, and arbitrary $\beta$, the thermodynamics of the deformed theory  is equivalent to that of ordinary Yang-Mills at leading order in the large $N$ expansion. This thermal generalization of volume independence is depicted in 
Fig.\ref{fig:L-deform} and described in Section 2.  
At {\it finite}  $N$, thermal volume independence implies that  the phase  and thermal properties   of the deformed theory in the interval:
\be
{\rm volume~independence: } \quad L \gtrsim (\Lambda N)^{-1}
\label{vind}
 \ee
for a given $\beta$ must coincide with the  finite temperature properties of ordinary Yang-Mills theory up to $O(1/N^2)$ corrections.  In the regime (\ref{vind}) the deformed theory remains  incalculable (without using lattice simulations) and the deconfinement transition cannot be studied analytically. However at smaller $L$, a calculable, semi-classical regime opens up:
\be
{\rm semiclassical~domain: } \quad L \lesssim (\Lambda N)^{-1}.
\label{sc-d}
 \ee
In this interval and at $\beta=\infty$,  we have   analytic control over the three-dimensional 
long distance dynamics, and  a weakly coupled  (yet non-perturbative) semi-classical 
realization of confinement \cite{Shifman:2008ja, Unsal:2008ch, Poppitz:2009uq}.\footnote{Such calculable regimes of QCD are, of course, not new. An analogous situation occurs in  describing hadronic matter at high density, where asymptotic densities provide  a weak coupling (but again non-perturbative) calculable framework where expected features of  hadronic matter at lower density   is reproduced \cite{Schafer:1998ef, Alford:1998mk}. Our deformation is, of course, more abstract, but morally similar.} For finite $\beta$, 
the effective description of the thermal theory is a two dimensional system  with electric and magnetic perturbations, 
in which $\beta$ appears as a parameter. 
The confinement-deconfinement transition can be studied analytically within this 2d model by varying $\beta$,   for arbitrary rank Yang-Mills theory, in contradistinction  with  \cite{Aharony:2005bq}.\footnote{The thermodynamic limit is achieved without need for $N=\infty$ as our set-up is at infinite spatial volume.}
The transition is plausibly smoothly connected to the deconfinement transition of the fully four-dimensional Yang-Mills theory. \footnote{We emphasize that the existence of a semi-classical domain in the deformed Yang-Mills theory and the absence of such a domain in ordinary thermal Yang-Mills theory does not contradict thermal large-$N$  equivalence, and volume independence. The semi-classical regime is $LN\Lambda \lesssim  1$, i.e., the scaling regime $L \sim (N\Lambda)^{-1}$ as $N \rightarrow \infty$, while the domain of large-$N$ equivalences   is the regime $L = O(N^0)$ as $N\rightarrow \infty$. As $N$ is increased, the semi-classical domain shrinks to a narrow sliver and the thermal large-$N$ equivalence holds for any $L =O(N^0)$. 
}

In this sense we find that the deconfinement transition of four dimensional Yang-Mills can be studied via a two dimensional field theory with electric and magnetic (order and disorder) perturbations. This  two dimensional system has parallels to the statistical mechanical systems studied in \cite{Fradkin:1980th}. In this framework the microscopic mechanism behind the deconfinement transition is a competition between electric and magnetic objects in a manifest and calculable way. At higher temperatures the electric objects are more relevant resulting in deconfinement, while at low temperatures magnetic objects are more relevant, resulting in confinement.

The fact that the deconfinement transition is manifestly driven by a competition of electric and magnetic degrees of freedom we feel is the most interesting qualitative aspect of our work, and the one most prone to generalization to other four dimensional gauge theories, perhaps including QCD. In fact, the idea of deconfinement as a competition between electric versus magnetic objects has already been introduced into real world QCD by Liao and Shuryak  as a possible way to explain some of the most interesting features of the quark gluon plasma at RHIC, in particular its relatively low viscosity to  entropy ratio. 
 Within the context of a simple toy model \cite{Liao:2006ry, Liao:2008jg} (also see  \cite{Chernodub:2006gu}) Liao and Shuryak argue  that the viscosity to entropy ratio of a plasma of electric and magnetic excitations is minimized when the densities of magnetic and electric objects are comparable.

The organization of this paper is the following. In Section 2 we elaborate on thermal large $N$ volume independence and introduce the deformed Yang-Mills theory. In Section 3 we discuss how the deformed Yang-Mills theory experiences weakly coupled confinement on $\R^3 \times \S^1_L$ at small $L$. In Section 4 we describe the calculable deconfinement transition and the role of electric and magnetic objects. In Section 5 we conclude and discuss novel directions suggested by this work.  

\section{Thermal large-$N$ equivalence}
\label{thermaleq}
Consider ordinary four-dimensional Yang-Mills theory compactified on $\R^2 \times \T^2$, parameterized as: 
\begin{equation}
{\cal M}_4(L, \beta) = \R^2 \times \S^1_L \times \S^1_\beta \; , 
\end{equation}
where $\S^1_\beta$ is a thermal circle of size $\beta$ while  $\S^1_L$ is an ordinary circle of radius $L$. The action is:
\begin{equation}
    S^{\rm YM}= \int_{ {\cal M}_4 } 
\left[    \frac{1}{2g^2} \>
    \tr\,  F_{\mu \nu}^2 (x)   + i \theta   \frac{1}{16 \pi^2}    
    \tr\,  F_{\mu \nu} \widetilde F^{\mu \nu}   \right]
    \,,
\label{eq:cont}
\end{equation}
where  $F_{\mu \nu}= F_{\mu\nu}^a t^a$ is non-Abelian field strength,  $ \widetilde F^{\mu \nu}  = \half \epsilon^{\mu\nu \rho \sigma} F_{\rho \sigma}$, $g$ is 4d gauge coupling, and $\theta$ is the theta angle.\footnote{ We normalize the generators $t^a$  of the Lie algebra in the defining  representation as:
 %
$ \tr (t^a t^b ) = \half \delta^{ab}\; .$
} For simplicity we henceforth set the $\theta$-angle to zero.

This theory possesses a global $(\Z_N)_\beta \times (\Z_N)_L$ center symmetry. This symmetry is the set of local $SU(N)$ rotations periodic up to an element of the center group of $SU(N)$: 
\begin{eqnarray} 
&& g(x_1, x_2, x_3 + \beta, x_4) = z_\beta  \; g(x_1, x_2, x_3 , x_4) ,  \cr
&& g(x_1, x_2, x_3, x_4 + L) = z_L  \;  g(x_1, x_2, x_3 , x_4),    \qquad  z_\beta^N = z_L^N=1.
\end{eqnarray}
moded out by the set of local gauge rotations (which are by definition single-valued on $\S_L^1\times \S_\beta^1$). The order parameters for the  $(\Z_N)_\beta \times (\Z_N)_L$ center symmetry are the  non-local Wilson lines:
\begin{equation}
\Omega_L = {\rm exp}[i \int_{\S^1_L} A_4 dx_4],  \quad \Omega_\beta= {\rm exp}[i \int_{\S^1_\beta} A_3 dx_3]
\end{equation}
along the $\S^1_L \times \S^1_\beta $ circles, respectively. The center symmetry acts on the order parameters as: 
   \begin{eqnarray}
 (\Z_N)_\beta \times (\Z_N)_L :&&\tr \Omega_\beta  \rightarrow z_\beta \;   \tr \Omega_\beta    \nonumber \\
 :&& \tr \Omega_L \rightarrow z_L \;   \tr \Omega_L  
\end{eqnarray}
We define the  \textbf{``$\Omega_L$-deformed Yang-Mills"} theory or simply \textbf{"deformed Yang-Mills"} as: 
\begin{equation}
    S^{\rm dYM} = S^{\rm YM} + \Delta S \,, \;\;
      \Delta S
    \equiv
    \int_{{\cal M}_4  }  \frac{1}{L^4} \>
     \sum_{n=1}^{\lfloor {N/2} \rfloor}
    a_n \left|\tr \left( \Omega^n_L \right) \right|^2 \,,
    \label{dYM}
\end{equation}\
with sufficiently positive coefficients $\{ a_n \}$ and $\lfloor N/2 \rfloor$ denoting the integer part of $N/2$.\footnote{
 Double-trace operators  are also   used in the works of Ogilvie et.al. to study phases with partial center symmetry breaking \cite{Ogilvie:2007tj, Meisinger:2009ne, Myers:2007vc}. Pisarski and collaborators give a phenomenological effective theory description of deconfinement by using such deformations \cite{ Pisarski:2006hz}, and study various aspects in \cite{Hidaka:2009ma}.  }

In the decompactifcation limit $\beta \rightarrow \infty$, the deformed theory enjoys volume independence \cite{Unsal:2008ch}:

\begin{quote}
{\bf Large-$N$ volume independence:} Yang-Mills theory on $\R^4$ is equivalent to the deformed Yang-Mills theory on $\R^3\times \S_L^1$ for any finite value of $L$, up to $1/N^2$ corrections, provided the $(\Z_N)_L$ center symmetry is not spontaneously broken.
\end{quote} 
Our deformed theory satisfies the condition that $(\Z_N)_L$ remain unbroken by construction. Equivalent means that correlation functions of \emph{neutral sector} observables - operators which are neutral under the $(\Z_N)_L$ center symmetry - are the same in the two theories up to $O(1/N^2)$ corrections.  
Volume independence does not apply to correlators containing non-neutral sector observables, the simplest example of which is  $\tr \Omega_L$.
Since  many interesting physical observables are in the neutral sector, many interesting observables in pure YM theory can be extracted by studying the correlators in  the deformed and  volume reduced theory.

The large-$N$ volume independence theorem has an immediate generalization to general $\beta \in (0,\infty)$:

\begin{quote}
{\bf Thermal large-$N$  equivalence:} Yang-Mills theory on $\R^3\times \S_\beta^1$ is equivalent to deformed Yang-Mills theory on $\R^{3-k}\times (\S_L^1)^k \times \S_\beta^1$ for any finite value of $L$ and for a given   $\beta$,   up to $1/N^2$ corrections, provided that the $[(\Z_N)_L]^k$ center symmetry is not spontaneously broken. 
\end{quote} 
As before, in our deformed theory  where $k=1$  the condition that $(\Z_N)_L$ remain unbroken is satisfied by construction. Also as before, the equivalence only applies to correlation functions of neutral sector observables. This important corollary to the volume independence theorem can be proven by a simple modification of the arguments of \cite{Unsal:2008ch}. Instead of discussing the proof, we will state its main physical implications.

Perhaps the most important part of the thermal equivalence is that the $(\Z_N)_\beta$-center symmetry is a {\it spectator symmetry}. It is left intact during the projections and deformations which are used in the chain of equivalences that are used to prove the volume independence theorem. Thus, the expectation values and connected correlators of topologically non-trivial Polyakov loops (which are neutral under $(\Z_N)_L$, but charged under $(\Z_N)_\beta$) are also part of the neutral sector to which the  thermal large-$N$ equivalence  applies. Thus these observables must agree in $\Omega_L$-deformed Yang-Mills and pure YM theory at {\it any} $\beta$. This means, the thermodynamics of the two theories are part of their respective neutral sector dynamics.

Thus, at leading order in $N$, the thermal Polyakov loops must agree, both in the confined and deconfined phases:
 \begin{equation}
\langle \frac{\tr}{N}  \tr \Omega_\beta \rangle^{\rm dYM}  (L) =\langle \frac{\tr}{N}  \tr \Omega_\beta \rangle^{\rm YM} = \left\{ \begin{array}{cc}
 0, & \qquad \beta > \beta_c,  \;\;\;{\rm confined} \cr
 z_\beta, &\;\;\; \qquad \beta <  \beta_c   \;\;\; {\rm deconfined}  
  \end{array} \right.
\label{exact2}
\end{equation}
where $z_\beta$ is some $N$-th root of unity. Furthermore the deconfinement temperature and the latent heat associated with the phase transition must also agree:
\begin{equation}
\beta_c^{\rm dYM}= \beta_c^{\rm YM} \left[1+ O(\frac{1}{N^2})\right] \;, \; \; Q^{\rm dYM}_l=   {Q}^{\rm YM}_l \left[1+ O(\frac{1}{N^{2}})\right] 
\label{exact}
\end{equation}
These agreements hold in the strongly coupled $LN\Lambda \gg 1$ domain where volume independence applies. As a consequence of this volume independence, these quantities are independent of $L$ at leading order in $N$.\footnote{The matching of the deconfinement temperature is  numerically  tested in  lattice regularized theory  by simulating  Yang-Mills theory with heavy adjoint fermions, a theory which emulates deformed Yang-Mills, 
 on a $ [(1)^3]_L  \times  2_{\beta}$ \cite{Azeyanagi:2010ne}, and agrees with large-scale lattice studies  \cite{Lucini:2005vg}. 
  }  

\subsection{Why bother?}
A common criticism of large-$N$ volume independence and in particular of deformation equivalences is that it maps a strongly coupled gauge theory to another strongly coupled gauge theory, neither of which is analytically calculable. So, why bother? 

It is true that in the strict $N\rightarrow \infty$ limit, neither of the equivalent pairs seem to be any easier. However, as we shall explain in the next two sections, at finite-$N$, the same deformation serves to engineer a semi-classical domain at small $L$. This semi-classical domain is  continuously connected to the strongly coupled regime of the undeformed theory. Such a step usually cannot be achieved within the undeformed theory itself. What the deformation achieves is a generalization of Yang-Mills theory that depends smoothly on an extra-parameter.\footnote{Small-volume or reduced formulation are also useful numerically.  For example, QCD with adjoint fermions satisfies volume independence if one uses periodic boundary conditions for all fields \cite{Kovtun:2007py}. Simulations and studies of the reduced QCD and related models  appears in  recent works \cite{Bringoltz:2009kb, Hietanen:2009ex, Azeyanagi:2010ne, Catterall:2010gx, Vairinhos:2010ha}. }

In this way one obtains a new weakly coupled regime of locally four dimensional gauge theories. Based on our experience with our best understood examples of non-perturbative quantum field theory, it is often useful to understand the various weakly coupled domains before attempting to understand the theory at strong coupling. Furthermore, the extension of calculations to the border of their validity can sometimes yield interesting information about the physics in the incalculable coupled domain. 



\section{Weak coupling confinement} 

It is easily seen that in the regime $LN\Lambda \lesssim 1$, the resulting effective long distance theory is a three-dimensional theory which enjoys weakly coupled confinement for a wide range of $\beta$. We consider first the limit $\beta \rightarrow \infty$ which has been studied in \cite{Unsal:2008ch}.  We briefly review their derivations to establish the context and notations for the next section.

In the zero temperature, weakly coupled domain, the deformed theory has a unique center-symmetric minimum for the Wilson line, $\Omega_L$. The fourth component of gauge field $A_4$ behaves as a compact adjoint Higgs field, and the theory reduces to a 3d Yang-Mills-Higgs system.  The vacuum expectation value of the Wilson line is 
   \begin{equation}
   \Omega_L= \eta~ {\rm Diag} \left( 1, e^{2\pi i/N},  e^{4 \pi i/N},  \ldots,  e^{2\pi i(N-1)/N} \right),
   \label{vev}
\end{equation}
where $\eta= e^{\pi i/N}$  for even $N$ and  $\eta=1$ otherwise, up to conjugation by gauge rotations. This leads to Abelianization (or adjoint Higgsing):  
\begin{equation}
SU(N) \rightarrow U(1)^{N-1},
\end{equation}
of the long distance dynamics. Since the fluctuations of eigenvalues are small due to the weak 
 't Hooft coupling, the Abelianization holds quantum mechanically.

Due to gauge symmetry breaking,  the off-diagonal components of the gauge field acquire masses. The spectrum of the gauge fluctuations in perturbation theory is composed of levels each of which is $N$-fold degenerate. The level spacing is $\frac{2 \pi }{LN}$. The masses and charges of the lightest $W$-bosons  are
\begin{equation}
 m_{W^{}_i}= \frac{2\pi}{LN},  
 \qquad Q_{W_i^{}} = g  \alpha_i     ,  \qquad i=1, \ldots,N.
\end{equation}
 Here, $\alpha_i   \in   \Delta^{0}_{\rm aff} -\{\alpha_N\}$ are the simple roots of the Lie algebra and $\alpha_N= -\sum_{i=1}^{N-1} \alpha_i$ is the affine root (which is there due to compactness of the adjoint Higgs). $\Delta^{0}_{\rm aff}$ is called the affine (extended) root system of the the associated Lie algebra,
\begin{equation}
    \Delta^{0}_{\rm aff} \equiv
    \{ \alpha_1, \alpha_2, \ldots , \alpha_{N-1},  \alpha_N \} \,.
\end{equation}
The roots $\alpha_i \in \Delta^{0}_{\rm aff}$ obey:\footnote{We changed the normalization with respect to  Ref.\cite{Unsal:2008ch} for convenience, such that the simple roots normalize to unity.}
%
\begin{equation}
    \alpha_i \cdot \alpha_j
    =  \delta_{i,j}
    - \half \delta_{i, j + 1}
    - \half   \delta_{i, j - 1}
    \,, \qquad i, j=1, \ldots N \,.
\label{Eq:inner}
\end{equation}

Since the gauge symmetry is broken as  $SU(N) \rightarrow U(1)^{N-1}$  due to the compact Wilson line
(\ref{vev}), there are   $N$ species of  monopole-instantons. 
The topological and magnetic quantum numbers of these instantons  are: 
 \begin{equation}
 \left( \int_{S^2} 
  \bm F,   \int \frac{1}{16 \pi^2}    
    \tr\,  F_{\mu \nu} \widetilde F^{\mu \nu}   \right) =  \left( \frac {4 \pi}{g} \,\alpha_i, \frac{1}{N} \right)
 \qquad
 \qquad
\label{eq:mag-charge}
\end{equation}
and the negation of those for the anti-instantons.\footnote{ These  monopoles are finite action topological configurations in the Euclidean formulation, and hence instantons \cite{Polyakov:1976fu} . $N-1$ of them are ordinary 3d instantons, and the extra instanton, which has no counterpart in a microscopically  3d theory and which is pertinent to locally 4d nature of the theory, is sometimes called a twisted instanton. In a center symmetric background, all of these instantons carry equal action.  Sometimes, they are also referred to as BPS and KK monopoles, or monopole-instantons.  Some results about Bogomolny-Prasad-Sommerfield (self-duality) equations, and relation between these 3d and 4d instantons are reviewed in the Appendix.\ref{chiral}.}

In three dimensions, Abelian duality relates a photon to a compact scalar. With $\sigma^j(\x) $ the compact scalar dual to the photon $A_{\mu}^j(\x)$ of the $j$-th  $U(1)$ subgroup, the Abelian duality relation is:
 \begin{equation}
	\bm F_{\mu \nu}  = \frac{g^2} {4 \pi  L} \>
	\epsilon_{\mu \nu \rho} \, \partial_{\rho} \bm \sigma \,.
    \end{equation}
To all orders in perturbation theory, (ignoring topological sectors), the long distance description  is free Maxwell theory in 3d, and is  given by: 
    \begin{equation}
	 S^{\rm pert.th.}
    =
    \int_{\R^3} \;
\frac{L}{4g^2} (\bm F_{\mu \nu})^2
	=     \int_{\R^3} \;
	\frac{1}{2L}
	\left(\frac{g}{4\pi}\right)^2
	(\nabla \bm \sigma)^2
	\label{free}
	    \end{equation}
The proliferation of these instantons leads to interaction terms in the Lagrangian of the compact scalar  \cite{Polyakov:1976fu}. This is the generalization of Polyakov's mechanism to a locally four dimensional gauge theory \cite{Unsal:2008ch}. The action for the  low energy effective theory (the dual Lagrangian) in the small   $\S^1_L$ domain is 
   \begin{equation}
    S^{\rm dual}
    =
    \int_{\R^3} \;
    \Big[
	\frac{1}{2L}
	\left(\frac{g}{4\pi}\right)^2
	(\nabla \bm \sigma)^2
	-
	\zeta \>
	\sum_{i=1}^N \>
	\cos ( \alpha_i  \cdot \bm \sigma) + \cdots
    \Big] \,,
\label{eq:Sdual}
\end{equation}
where $    \zeta \equiv
    C \, e^{-S_0}
    = A \,  m_W^3 \, (g^2N)^{-2} \, \, e^{-8\pi^2/ (g^2N (m_W))}$ is the monopole fugacity, and $S_0$ is the instanton action.

The action (\ref{eq:Sdual}) is a non-renormalizable low energy effective theory valid at distances 
larger than $ m_W^{-1} \sim LN$. Ellipsis stands for higher order terms in the semi-classical expansion as well as terms due to the omission of $W$-bosons.

The existence of a mass gap and linear confinement can easily be derived using the dual Lagrangian (\ref{eq:Sdual}). The mass gap for the $(N-1)$ photon species is:
\begin{equation}
m_p =m_\sigma  \; {\rm sin} \left(\frac{\pi p}{N} \right),  \qquad 
m_\sigma=A\Lambda  \left( \frac{LN\Lambda}{2 \pi} \right)^{5/6}  \left|{\rm ln} \left( \frac{LN\Lambda}{2 \pi} \right) \right|^{8/11}, \qquad p=1, \ldots, N-1
\label{massgap}
\end{equation}
where $A$ is an $O(1)$ coefficient.  The result of the semi-classical analysis is reliable in the 
\begin{equation}
  \frac{LN\Lambda}{2 \pi}  \ll 1
\end{equation}
domain.\footnote{The power of logarithm,  given in   given in Eq.(3.37) of \cite{Unsal:2008ch} as $9/11$,  is a minor error.  More importantly, the small parameter in the 
the   discussion of Ref.\cite{Unsal:2008ch}  is actually  $\left( \frac{LN\Lambda}{2 \pi} \right)$, not 
$LN\Lambda$. Although the former is manifest in the formulae, the factor of $2 \pi$ was  
not explicitly written. It is  actually  useful to restore it. }
 Obviously, the precise quantitative features of the mass spectrum and the string tensions  
have a non-trivial  $L$ dependence in the semi-classical domain. At strong coupling $LN\Lambda \gg 2 \pi$ and  at leading order in the large-$N$ expansion, all the neutral sector observables must saturate to constants independent of $L$ due to volume independence (see Section \ref{thermaleq}).  Naturally, one expects the semi-classic description to match the strong coupling description around   $LN\Lambda \sim 1$. 


\section{Calculable deconfinement}
\label{CD}
The deformed Yang-Mills theory on  $ \S^1_L \times \R^3 $ exhibits weakly coupled 
confinement  in the semi-classical domain ($LN\Lambda \lesssim 1$), where the theory experiences 
adjoint Higgsing. This ``Higgsed" regime is analytically connected to the  $LN\Lambda \gg 1$  regime and to  the theory on $\R^4$ in the sense that  there exist no order parameters which can distinguish the two-regimes.  In this section, we develop a formalism in the semi-classical domain which permits us to  study  the thermal phase transition. To do so, we consider a 
 finite temperature compactification of the deformed theory on $ \S^1_L \times \R^3 $  which corresponds to  the theory on $ {\cal M}_4(L, \beta) =\S^1_L \times \R^2 \times \S^1_\beta $ at arbitrary $\beta$.  \footnote{The analysis of the thermodynamics of the deformed Yang-Mills theory is  analogous to the one of the finite temperature 3d Georgi-Glashow model \cite{Agasian:1997wv, Dunne:2000vp, Kogan:2001vz, Kovchegov:2002vi} however, our set-up differs  from it in the sense of being locally four dimensional. }


First, let us momentarily ignore $W$-bosons. (This assumption and its region of validity will be examined below.) 
 At asymptotically low temperatures, $\beta  \gg m_\sigma^{-1}$,  the dynamics is that of the 3d dual theory  (\ref{eq:Sdual}). A more interesting regime is 
\begin{equation} 
m_W^{-1} \ll \beta \ll m_\sigma^{-1} \sim m_W^{-1} e^{S_0/2} \; , 
\end{equation}
where  the size of the monopoles is much smaller than $\beta$ which in turn  is much smaller than the inter-monopole separation. In this regime the potential induced by  a monopole which is $1/r$ in 3d is enhanced 
to $\log(r)$ at large distances which is the Coulomb potential of a charge in 2d. This can be seen by using the  method of images from electrostatics. 
To incorporate this effect in field theory, it suffices to compactify the low energy effective theory (\ref{eq:Sdual}) down to 2d. 
In this domain, the  theory reduces  to a well-known two dimensional theory of  ``vortices", which are the dimensional reduction of 3d instantons. The action is:
\begin{equation}
    S^{\rm dual}
    =
    \int_{\R^2} 
    \Big[ \frac{a}{2}
	(\nabla  \bm \sigma)^2
	-
	\zeta_M \>
	\sum_{i=1}^N \>
	\cos (\alpha_i \cdot \bm \sigma) + \cdots
    \Big] \, , \;\; a \equiv \frac{\beta}{L}
	\left(\frac{g}{4\pi}\right)^2
\label{eq:Sdual2}
\end{equation}
where   $\zeta_M= \beta \zeta$.  For $N=2$, this is the sine-Gordon model in $d=2$ dimensions, and it is its generalization  for $N\geq 3$.   
 Whether a mass gap for the $\bm \sigma$ field is generated or not is tied with the question of the relevance of the  $ e^{i  \alpha_i \cdot \bm \sigma} $ operator. 
The conformal dimension of the operator  $ e^{i   \alpha_i \cdot \bm \sigma} $ about the free scalar fixed point is:
\begin{equation}
\Delta [e^{i   \alpha_i \cdot \bm  \sigma}]= \frac{   \alpha_i^2   }{4\pi a} =  
\frac{1 }{4\pi a}  =  \frac{4 \pi L}{\beta g^2} ; 
  ,  
\label{dim1} 
\end{equation}
Note that for all $\alpha_i   \in   \Delta^{0}_{\rm aff}$, the conformal dimensions are identical, because the algebra is simply-laced, $\alpha_i^2 =1, i=1, \ldots, N$.\footnote{ \label{foot:rel} The ellipsis in 
(\ref{eq:Sdual2}) stand for perturbations sub-leading in the semi-classical expansion. Here, there are some subtle issues. 
 Even at order $k$ in the expansion, there is a sub-class of operators which has the same scaling dimension as the leading term, for example, $ \Delta [  e^{- k \frac{8 \pi^2}{g^2N}}  e^{i (\alpha_i + \alpha_{i+1}  + \ldots \alpha_{i+k})  \cdot   \bm { \sigma} }] = 
  \Delta [   e^{-\frac{8 \pi^2}{g^2N}}  e^{i \alpha_i \cdot   \bm {\sigma} }]  $ due to a Lie algebra identity, $(\alpha_i + \alpha_{i+1}  + \ldots \alpha_{i+k})^2=1$ for $k \neq N$.  In the effective theory, these and a plethora of   others      are there, generated and relevant in the sense of 
Wilsonian renormalization group.  Although the scaling dimensions of these operators are identical to the ones that appeared in our 
effective Lagrangian (\ref{eq:Sdual2}), in the weak coupling domain, their prefactors are suppressed by extra-powers of  $ e^{-\frac{8 \pi^2}{g^2N}}$. Hence, 
they remain as small perturbations at distances where the leading magnetic perturbation becomes strong. Thus, the effect of sub-leading  terms are negligible there. Close to the boundary of semi-classical window, these operators may and will become important, as well as possibly near the critical temperature for the deconfinement transition.}

The perturbation of the free theory by the vortices is relevant  if the conformal dimension 
$\Delta$ is less than two, irrelevant for $\Delta$ greater than two, and marginal otherwise. 
The  quantum theory  of  (\ref{eq:Sdual2})  undergoes a phase transition at 
$\Delta= 2$, 
\begin{equation}
\beta_m = \frac{2 \pi L}{g^2}
\label{bm}
\end{equation}
where  subscript $m$ stands for magnetic, between a phase of finite correlation length at {\it low} temperatures  $\Delta <2$ ($\beta > \beta_m $),  and a phase of infinite correlation length at {\it high} temperatures $\Delta >2$ ($\beta < \beta_m$).  
This is the well-known Berezinsky-Kosterlitz-Thouless   (BKT) 
transition \cite{Berezinsky:1970fr, Kosterlitz:1973xp}, albeit 
with an {\it inverted} temperature. In other words,  the high  temperature phase is 
 populated by   neutral magnetic vortex-anti-vortex pairs  and these pairs dissociate at low temperature, opposite to the conventional BKT transition.  This means,  in the low and zero  temperature phase, the mass gap is induced by the magnetic  defects in the $\Omega_L$-deformed  Yang-Mills  theory.  
  
 However, the effect described above is not the whole picture - the gapless phase is an artifact  associated with the omission of electrically charged W-bosons, as noted in the context of the 3d Georgi-Glashow model in \cite{Dunne:2000vp, Kogan:2001vz}.  The W-bosons are not important in the long distance regime of the gauge theory on  $\R^3 \times \S^1_L$ because they are finite energy (mass) particles, as opposed to 3d instantons which are finite action defects. However, when the space is further compactified to  $\R^2 \times \S^1_\beta \times \S^1_L$, W-bosons traveling around the thermal $\S^1_\beta$ circle have finite action, equal to 
 $\beta m_W$.   Their Boltzman weight is 
 $e^{-\beta m_W} $ and has an interpretation as a W-boson fugacity. The  W-bosons  are  a small perturbation  (with respect to topological defects) when     $e^{-\beta m_W}  \ll e^{-S_0}$ or $\beta \gg  \frac{4 \pi L}{g^2}$. Clearly, the scale at which monopoles become irrelevant is    
 outside this regime.

\begin{figure}[ht]
\begin{center}
\includegraphics[angle=-90, width=0.80\textwidth]{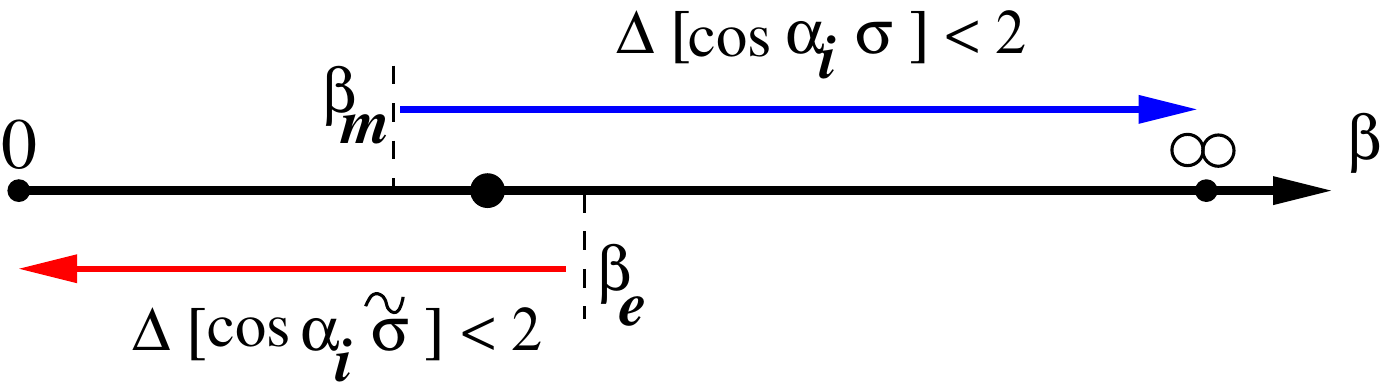}
\caption{
For  $\beta > \beta_m$, $e^{i \alpha_i \cdot \bm  \sigma}$ is relevant, and for   $\beta < \beta_e$, $e^{i \alpha_i \cdot \bm {\widetilde \sigma}}$ is relevant.    In  $\beta_m < \beta < \beta_e$ interval, both perturbations are relevant. 
}
  \label {fig:relevance}
\end{center}
\end{figure}

If one ignores the topological sectors of gauge theory, which is justified if  $e^{-\beta m_W}  \gg e^{-S_0}$  ($\beta \ll  \frac{4 \pi L}{g^2}$), the proliferation of the two-dimensional gas of $W$-bosons generates an effective theory 
\begin{equation}
    S^{\rm p.t.}
    =
    \int_{\R^2} 
    \Big[ \frac{\tilde a}{2}
	(\nabla  \bm {\widetilde \sigma})^2
	-
	 \zeta_W \>
	\sum_{i=1}^N \>
	\cos (  \alpha_i \cdot  \bm {\widetilde \sigma}) + \ldots
    \Big] \, , \;\; \tilde a \equiv \frac{1}{16 \pi^2 a} = \frac{L}{\beta g^2}
 \label{eq:Sdual3}
\end{equation}
where $\zeta_W \sim \frac{1}{\beta^2} e^{-\beta m_W}$ and   $\frac{1}{4 \pi a}  * d \bm {\widetilde \sigma}  = d \bm \sigma $ is the dual of $\bm \sigma$ field in 2d.   
The conformal dimension of the $W$-boson operator is 
\begin{equation}
\Delta [e^{i  \alpha_i \cdot   \bm {\widetilde \sigma} }]= \frac{\alpha_i^2 }{4\pi \tilde a} = 4\pi a =   \frac{\beta g^2}{4 \pi L} \;  , 
\label{dim2}
\end{equation}
The ellipsis in (\ref{eq:Sdual3}) stands for electric perturbations sub-leading  in 
$e^{-\beta m_W}$ expansion, and the analog of the discussion in footnote (\ref{foot:rel})  applies. 
This theory has a BKT transition at $\Delta [e^{i  \alpha_i \cdot   \bm {\widetilde \sigma} }]=2$ or:
\begin{equation}
\beta_e = \frac{8 \pi L}{g^2},
\label{be}
\end{equation}
 where subscript $e$ stands for electric. 
It  has a  gapped phase at high temperatures $\beta < \beta_e $ induced by free electrically charged excitations and a gapless phase at low temperatures where electrically charged excitations form neutral molecules. This makes sense because in the absence of topological defects, the large-$\beta$ theory is  the compactification of the free Maxwell theory  (\ref{free}) which is  related to the gapless phase of (\ref{eq:Sdual3}) via an $a\leftrightarrow 1/a$ or T-duality.

 The magnetic monopoles   are  a small perturbation  (with respect to $W$-bosons) when     $
  e^{-S_0}  \ll e^{-\beta m_W} $ or $\beta \ll  \frac{4 \pi L}{g^2}$. Clearly, the scale at which 
  $W$-bosons  become irrelevant is     outside this regime. 
This implies that neither electric nor magnetic  BKT  is actually there  while (\ref{eq:Sdual2}) and 
(\ref{eq:Sdual3}) are valid descriptions.  
 \footnote{
If, in  Fig.\ref{fig:relevance},   $\beta_m$ were larger than $\beta_e$ within the regions of validity of   (\ref{eq:Sdual2}) and (\ref{eq:Sdual3}), this would have implied the presence of two genuine BKT 
transitions with an intermediate gapless phase.  Indeed,  in the  planar  Heisenberg model of  ferromagnetism with a symmetry breaking perturbation (which reduces the symmetry of the theory to $\Z_N$),  such a phenomena takes place for all $N > 4$ \cite{Jose:1977gm}. }

At arbitrary $\beta$, and in particular, in a domain where both electric and magnetic perturbations are relevant,  we should instead consider a Lagrangian of the form: 
\begin{equation}
    S
    =
    \int_{\R^2} 
    \Big[ \frac{a}{2}
	(\nabla  \bm {\sigma})^2
	-  \zeta_M \>
	\sum_{i=1}^N \>
	\cos (  \alpha_i \cdot  \bm {\sigma})  
	-
	 \zeta_W \>
	\sum_{i=1}^N \>
	\cos (  \alpha_i \cdot  \bm {\widetilde \sigma}) + \ldots
    \Big] \, ,
 \label{eq:Sdual4}
\end{equation}
where in the path integral we have to impose the duality relation $\frac{1}{4 \pi a}  * d \bm {\widetilde \sigma}  = d \bm \sigma $ as a constraint. The electric-magnetic Coulomb gas representation associated with the field theory has the form $V_{\rm int} = V_{e-e} + V_{m-m} + V_{e-m} $ where $V_{e-e}$ ($V_{m-m}$)   is   the mutual logarithmic Coulomb interaction of electrically (magnetically) charged excitations  and  $V_{e-m}$  
is the interaction between electrically and magnetically charged excitations. For a detailed description and references to earlier related works,  we   recommend  the reader Ref.\cite{Kovchegov:2002vi}.


\subsection{Electric-magnetic competition and  relation to Polyakov order parameter}

In the deformed Yang-Mills theory, the confinement-deconfinement transition is explicitly realized as a competition between electric and magnetic perturbations. This is a calculable realization of the scenario proposed in Ref.\cite{Liao:2006ry}. 
There are  three regimes as a function of $\beta$, as shown in Fig.\ref{fig:relevance}. 
 For $\beta > \beta_e$,    
the $e^{i    \alpha_i \cdot  \bm \sigma}$ are relevant while the $e^{i     \alpha_i \cdot  \bm{  \widetilde \sigma } }$ are  irrelevant. In this phase, magnetic defects are free and dominate the long-distance dynamics, while the electrically charged particles are confined.  
 For $\beta < \beta_m$, the situation is reverted: 
the $e^{i     \alpha_i \cdot  \bm \sigma}$ are irrelevant while the $e^{i      \alpha_i \cdot  \bm{ \widetilde \sigma}}$ are relevant, which means that electric charges are free while magnetic defects are confined. 
 In the interval $ \beta_m < \beta < \beta_e $, both  $e^{i    \alpha_i \cdot  \bm  \sigma}$  and $e^{i     \alpha_i \cdot  \bm {\widetilde \sigma}}$  are relevant - we discuss this domain in more detail in Sec.\ref{PT}.

In the small $LN\Lambda$ domain, since the IR theory Abelianizes, the fundamental Polyakov loop may be  identified with a  `` fundamental Quark"-operator. 
We define the following mapping 
\begin{equation}
 \frac{\tr}{N} \Omega_{\beta}   \;\;  \leftrightsquigarrow \;\;    \frac{1}{N}  \sum_{i=1}^{N} e^{i   \nu_i \cdot  \bm {\widetilde \sigma} }     
\end{equation}
where  $\nu_i, i=1, \ldots, N $ are the weights associated with the electric charges of the quarks in the fundamental representation.   
\footnote{ {\bf Conventions}:  \label{conventions} The $N$ weights 
$\nu_i$ are $N-1$ dimensional vectors forming an $(N-1)$-simplex. They satisfy 
\begin{equation}
 \nu_i \cdot  \nu_j =  \half \left( \delta_{ij} - \frac{1}{N} \right),   \;\; i, j=1, \ldots, N  \; . 
\end{equation}
 $(N-1)$-simplex is  the figure associated with the defining representation of the algebra.  
 At this stage, it is also useful to define the fundamental weights $\mu_k$, 
\begin{equation} 
\mu_k = \sum_{j=1}^k \nu_j, \qquad k=1,\ldots, N-1
\end{equation}
Fundamental weights form the weight lattice 
$\Lambda_w$, and the simple roots form the dual root lattice $\Lambda_r$.  $\Lambda_r$ is a sub-lattice of  $\Lambda_w$ and the quotient is isomorphic to  $\Lambda_w/\Lambda_r=\Z_N$.  The generators of the $(\Lambda_w,\Lambda_r)$ obey 
 \begin{equation} 
\alpha_i \cdot \mu_j  = \half \alpha^2_i \delta_{ij}   = \half {\delta_{ij}}
 \qquad i, j=1,\ldots, N-1, 
\end{equation}
the reciprocity relation. }
If the external charge sourcing the Polyakov loop is in some other representation, the mapping generalizes straightforwardly.\footnote{ 
The  generalizations of this mapping to anti-symmetric,  symmetric  and adjoint 
 representations are: 
\begin{eqnarray}
&& \tr \Omega_{\beta, {\rm AS}}   \;\;  \leftrightsquigarrow \;\;      \sum_{i < j =1}^{N} 
  e^{i  ( \nu_i  +  \nu_j ) \cdot  \bm {\widetilde \sigma} }      \qquad 
  \tr \Omega_{\beta, {\rm S}}   \;\;  \leftrightsquigarrow \;\;      \sum_{i \leq j =1}^{N} 
  e^{i  ( \nu_i  +  \nu_j ) \cdot  \bm {\widetilde \sigma} }      \qquad 
  \tr  \Omega_{\beta, {\rm Adj}}   \;\;  \leftrightsquigarrow \;\;      \sum_{i, j=1}^{N}  
    e^{i   (\nu_i  - \nu_j ) \cdot  \bm {\widetilde \sigma} } \; . \qquad \; 
\end{eqnarray}  
}

The Lagrangian  (\ref{eq:Sdual3}) apart from the obvious 
periodicity  identification 
$ \bm {\widetilde \sigma} \sim  \bm {\widetilde \sigma} + 4 \pi \alpha_j$, is also invariant under 
a discrete $\Z_N$ which we identify with the   ordinary  center symmetry,  $(\Z_N)_\beta$.  
A  shift in the weight lattice $\Lambda_w$  acts as  
\begin{eqnarray}
(\Z_N)_\beta : \;&&  \; \bm {\widetilde \sigma} \rightarrow 
  \bm {\widetilde \sigma} - 4 \pi \mu_k  \cr 
 : \; &&  e^{i   \nu_i \cdot  \bm {\widetilde \sigma} }     \rightarrow 
 e^{ + i \frac{2 \pi}{N}k} \; 
  e^{i   \nu_i \cdot  \bm {\widetilde \sigma} }   
  \label{center2}
  \end{eqnarray}
In reaching the second step, we used the identities given in footnote.\ref{conventions}. 
Let us now calculate the the expectation value of the Polyakov loop.

\subsubsection{Low temperature}

In the $\beta \gg  \beta_e$ domain, we can  safely use the Lagrangian (\ref{eq:Sdual2}) to describe the dynamics.  The  electric perturbations are highly suppressed and also  irrelevant  in the renormalization group sense. 
The insertion of an electric charge into the medium  may be viewed as a vortex in the  
$\bm \sigma$ field  theory.  The vorticity is the electric charge associated with the probe, i.e., 
\begin{equation}
\frac{Q}{g}= \frac{1}{2\pi}  \int_C  d \bm \sigma = \nu_j
\end{equation}
where $C$ is a closed curve encircling the test charge. Thus, we have 
\begin{equation}
\bm \sigma (\theta) \sim     \nu_i  \theta, \qquad \nabla \bm \sigma   \sim \frac{\nu_i}{r}  .
\label{ansatz}
{\bf e_{\theta}}
 \end{equation}

To evaluate  the action of a vortex in the free theory, we regularize the $\R^2$ space to  a disk $D_2(R)$ with radius $R$.  (We also need a short distance cut-off.  The finite size of vortex core serves this goal,   but this short-distance divergence is unimportant 
 for what follows.)  It is: 
\begin{equation}
S(R) = \int_{D_2(R)} d^2x \;  (\nabla \bm \sigma)^2   \sim \int^R  r dr   \frac{1}{r^2}  \sim \log R
\end{equation}
This is, indeed, the Coulomb potential of a test charge in 2d and it  clearly diverges 
as $R \rightarrow \infty$. When we take into account the potential 
 (\ref{eq:Sdual2}),  we observe that  the action   grows  quadratically: 
 $S_{int. }(R) \sim  \int_{D_2(R)} d^2x \cos  (\alpha_i \cdot \bm \sigma)  \sim R^2$.  However, this is an overestimation due to the form of the ansatz (\ref{ansatz}). The minimization of action in the space of possible  $\bm \sigma (\theta)$ with the given vorticity generates  a linearly rising action as $R$ is increased.   This is a configuration where ${\bm \sigma}$ is constant everywhere, but exhibits a jump along a  cut. The punch-line is, in the confined phase, we have 
 \begin{equation}
\langle  \frac{1}{N}  \sum_{i=1}^{N}  e^{i   \nu_i \cdot  \bm {\widetilde \sigma} }   \rangle =
 \lim_{R \rightarrow \infty} \left[  \frac{1}{N}  \sum_{i=1}^{N}   \left( e^{-  S [ {\rm in \; the \;  presence \;  of  \; } \frac{1}{2\pi}  \int_C  d \bm \sigma = \nu_j ] - 
S[ {\rm in \; its  \;  absence\;}] }  \right) \right]= 0
\end{equation} 

\subsubsection{High temperature}

In the $\beta \ll  \beta_m$ domain, we can   use the Lagrangian (\ref{eq:Sdual3}) reliably. 
The magnetic excitations are suppressed and irrelevant. 
  Here we wish to calculate the expectation value of the  $e^{i   \nu_i \cdot  \bm {\widetilde \sigma} }  $ operator in a description where  $ \bm {\widetilde \sigma} $ is the local field describing Lagrangian. 
 The periodicity identification of the $\bm {\widetilde \sigma}$ field is 
$ \bm {\widetilde \sigma} \sim  \bm {\widetilde \sigma} + 4 \pi \alpha_j$.  The potential 
$V (   \bm {\widetilde \sigma})  = -  \zeta_W \>
	\sum_{i=1}^N \>
	\cos (  \alpha_i \cdot  \bm {\widetilde \sigma})  $ is also invariant under the $(\Z_N)_\beta$ 
	 center symmetry (\ref{center2}), and has $N$  isolated minima within the unit-cell of the root lattice. This means that the theory has $N$  thermal equilibrium states in this phase. The expectation value of Polyakov loop is: 
\begin{equation}
 \langle  \frac{1}{N}  \sum_{i=1}^{N} e^{i   \nu_i \cdot  \bm {\widetilde \sigma} }     \rangle =  e^{ + i \frac{2 \pi}{N}k} \; ,  \qquad  k=0, 1, \ldots, N-1 \; . 
 \end{equation}
  These $N$-minima can be rotated into each other by the action of  spontaneously broken $(\Z_N)_\beta $ symmetry.

This is precisely the picture that we believe should hold in Yang-Mills theory. In deformed Yang-Mills, we analytically demonstrated the existence of the two phases.

\vspace{3mm}

\vspace{3mm}

\noindent  \emph {Remark:} The main novelty of this  description is following: The effective dual Lagrangians,  (\ref{eq:Sdual}), (\ref{eq:Sdual2}) and (\ref{eq:Sdual3}), are already long-distance descriptions. The non-perturbative phenomena, such as a mass gap, linear confinement in the confined phase, and the existence of a deconfinement transition, are already in the tree-level description of the dual theory. This is a main difference between studies of deconfinement to date and our description. In our dual formulation, the long-ranged correlations are already built into the dual Lagrangians and correlation functions can be easily evaluated  via these actions. This progress is  possible because toroidal compactification with deformation introduces a new parameter, $LN\Lambda$, in the theory. When this parameter is taken large, we face the conventional problems of strong gauge  dynamics.  

\subsection{Estimate for phase transition scale} 
\label{PT}
In the semi-classical domain, the theory has at least two phases,   $\beta<\beta_m$ where electric charges are free and magnetic charges are confined, and $\beta>\beta_e$ where magnetic charges are free and electric charges are confined. The phase transition must occur at some:
\begin{equation}
\beta_c  \in [\beta_m, \beta_e]= \Big[\frac{2\pi L}{g^2}, \;   \frac{8\pi L}{g^2}\Big] \;. 
\label{interval}
\end{equation}
In this domain, we do not have a good tool to find the value of the transition temperature, as both 
perturbations are relevant.  

Despite the fact that we can demonstrate the existence of two phases (confined and deconfined) in a semi-classical approximation,  the transition itself takes place in a regime (\ref{interval}) where the theory again becomes strongly coupled!

We  conjecture that the transition should occur when both  electric and magnetic perturbations simultaneously become order one following \cite{Kovchegov:2002vi}. To argue this, note that if one perturbation is order one while the other is small,  then the system is gapped  either due to electric excitations or magnetic excitations, which is to say the system is in one of the two phases.   In such a case, the smaller effect  may be treated within non-degenerate perturbation theory, and should not alter the behavior of the theory drastically. 
The two types of perturbations become comparable when the densities (i.e., fugacities) of electrically and magnetically charged quasi-particles become comparable. This is also argued to be the case in the scenario of Ref.\cite{Liao:2006ry} within the context of QCD. 
Indeed,  for $SU(N)$, 
     $e^{-\beta m_W}= e^{- \beta \frac{2\pi}{LN}}  \sim  e^{-S_0}= e^{-\frac{8 \pi^2}{g^2N}}$ 
at:
\begin{equation}
\beta_c = 
 \frac{4 \pi L}{g^2}=   \frac{4 \pi LN}{\lambda} \; ,
 \label{critical0}
 \end{equation}
 which is actually the midpoint of the (\ref{interval}). 
 
 It is also instructive to study the conformal dimensions of the electric and magnetic   perturbations  around the deconfinement temperature.  We find: 
 \begin{eqnarray}
&&\Delta_e \equiv \Delta [e^{i  \alpha_i \cdot   \bm {\widetilde \sigma} }] 
=  \frac{ Q_{W_i}^2 \beta }{4\pi L} =  \frac{ g^2  \beta }{4\pi L}   ,  \quad 
\Delta_{m} \equiv  \Delta [e^{i  \alpha_i \cdot   \bm { \sigma} }]=   \frac{ Q_{M_i}^2 L }{4\pi \beta}   
=  \frac{4\pi L} { g^2  \beta }
  \cr  \cr
 &&  \Delta_e  \Delta_{m} = \frac{(Q_{W_i} Q_{M_i})^2}{(4\pi)^2} =1  \qquad [ \rm no \; sum  \; over  \; i ]
  \end{eqnarray}
  where the reciprocity of the dimensions of electric and magnetic perturbations is a consequence of the Dirac quantization condition 
  \begin{equation}
  Q_{W_i} \cdot Q_{M_j} = g \alpha_i \cdot \frac{4 \pi}{g} \alpha_j = 4\pi \left(\delta_{ij}- \half \delta_{i,j+1}
 - \half \delta_{i,j-1} \right)
  \end{equation} 
At the critical point, the  dimensions of both  perturbations are equal to one, 
  \begin{equation}
\Delta [e^{i  \alpha_i \cdot   \bm {\widetilde \sigma} }] \Big |_{\beta_c} = \Delta [e^{i  \alpha_i \cdot   \bm {\sigma} }] \Big |_{\beta_c} =1 \; .
\end{equation}
\begin{figure}[ht]
\begin{center}
\includegraphics[angle=-90, width=0.80\textwidth]{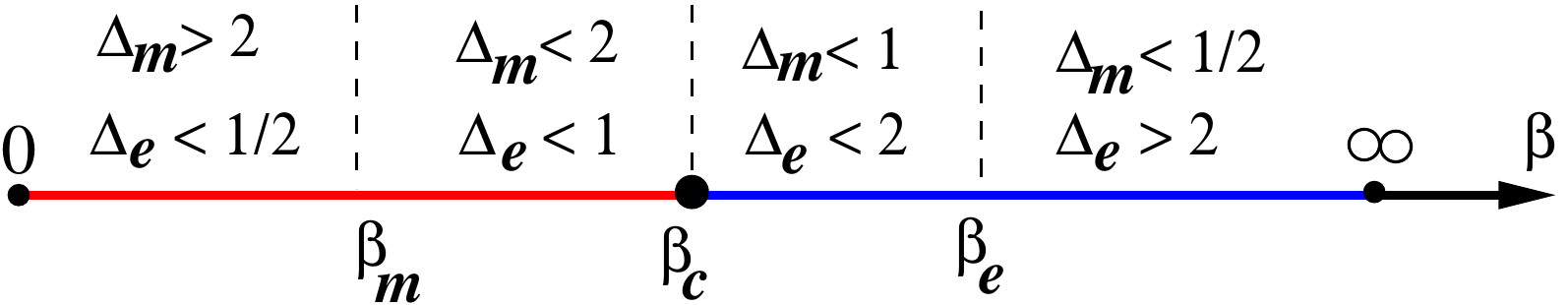}
\caption{
A  more refined  version of  Fig.2. 
 The putative phase transition is expected to occur at 
 $\beta_c$ where  $\Delta_e= \Delta_m=1$.   For  $ \beta_c  < \beta < \beta_e$,  magnetic operators dominate, and 
 $ \beta_m  < \beta < \beta_c$, electric  operators dominate. 
}
  \label {fig:refined}
\end{center}
\end{figure}
In this sense, the theory as a function of $\beta$ has four interesting domains and plausibly a single phase transition at $\beta_c$, as depicted in Fig.\ref{fig:refined}.
The theory exhibits confinement for  $\beta> \beta_c$, which corresponds  to $\Delta_m  <1, \;  \Delta_e >1$ and deconfinement for $\beta< \beta_c$ which corresponds to $\Delta_m >1,  \; \Delta_e<1$.   

This provides a more refined version of the domains of the thermal gauge theory relative to the Polyakov order parameter. We will speculate on the possible significance of $\beta_m$ in the conclusions.

\subsection{Extrapolation to larger $L$ or larger $N$}

\begin{figure}[ht]
\begin{center}
\includegraphics[angle=0, width=0.50\textwidth]
{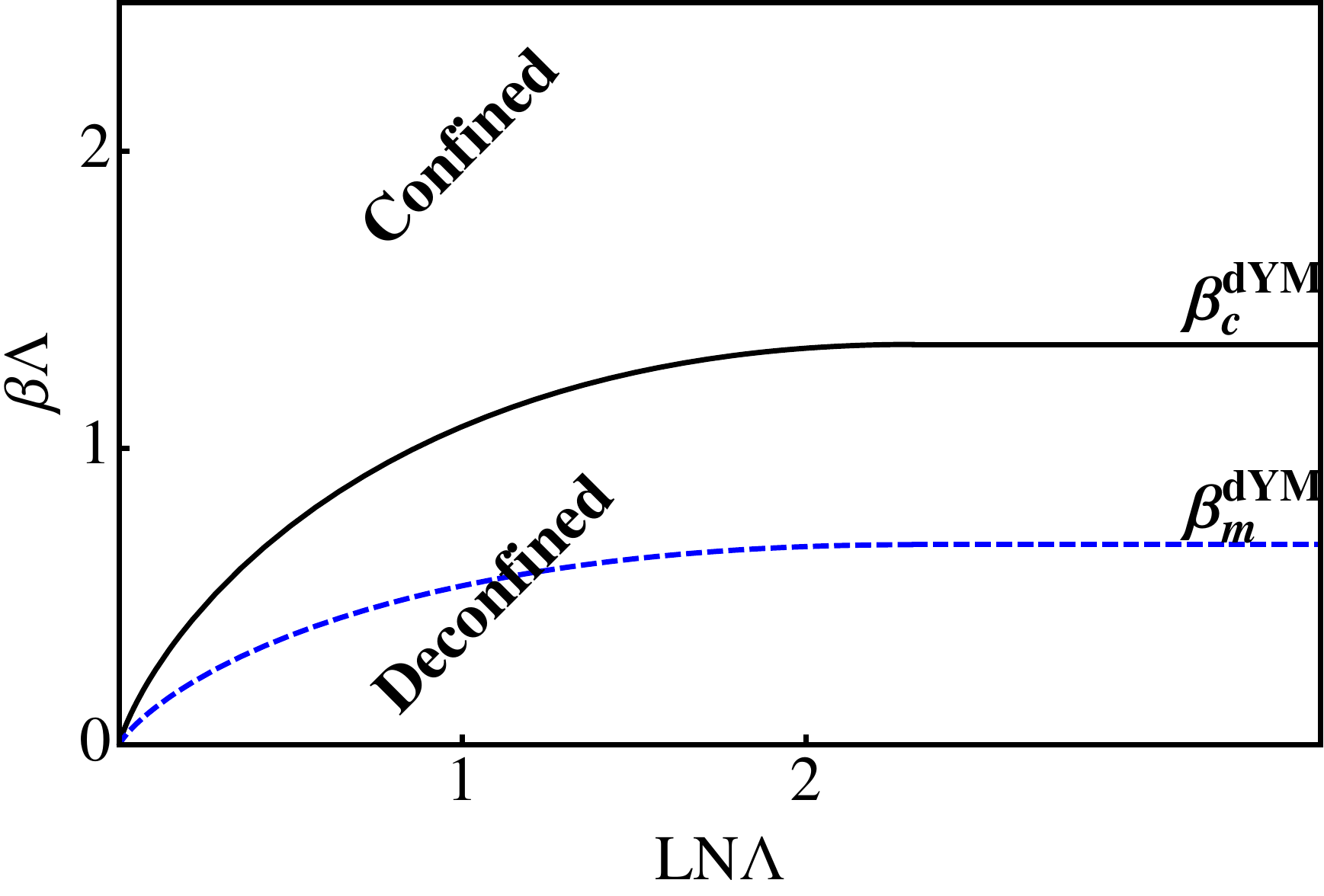}
\caption{
Simplest possible  phase diagram of $SU(N)$  deformed Yang-Mills  theory  on  $\R^2 \times \S^1_\beta \times \S^1_L$.   Above (below) the solid line, the theory is in the  $(\Z_N)_\beta$ unbroken  (broken) confined (deconfined) phase. 
Between the solid and dashed line $T_c <T<2T_c$, and at least in the  semi-classical domain, although the theory is in deconfined phase, magnetic defects  are still relevant. Below the dashed line, they  are  irrelevant. 
}
  \label{fig:phase}
\end{center}
\end{figure}

The appearance of $N$ in (\ref{critical0})  is   rather crucial, because the region of validity of the semi-classical analysis  is  $LN\Lambda \lesssim  1$,  not  $L\Lambda \lesssim 1$ (otherwise this would clash 
 with large-$N$ volume independence).   At the boundary of the semi-classical domain, the transition temperature  approaches $\beta_c  \sim \Lambda^{-1}$, the expected result based on numerical simulations and dimensional analysis.   For   $LN\Lambda \gg 1$, by volume independence, this value must be  saturated, up to  $1/N^2$ corrections, hence the plateau shown in Fig.\ref{fig:phase}.
 The  critical temperature is, 
 $T_c= \beta_c^{-1}$:
 \begin{equation}
\beta_c^{\rm dYM} = \left\{\begin{array}{cc} 
\frac{11}{3 } \Lambda^{-1}  \left( \frac{LN\Lambda}{2 \pi} \right) \left|{\rm ln} \left( \frac{LN\Lambda}{2 \pi} \right) \right|,
&  \qquad  LN \Lambda \lesssim 1 \\
  c \Lambda^{-1}\left(1+ O(1/N^2)\right) & \qquad   L N \Lambda \gtrsim 1 \\
\end{array}
\right.
\label{critical}
\end{equation}
In order to reach to the volume independence domain, we do not necessarily need to increase $L$. We can keep $L$-fixed while  increasing $N$.  The base space remains {\it macroscopically} two dimensional, but the dynamics (and thermodynamics) of the theory interpolates to that of Yang-Mills theory on $\R^3 \times \S^1_\beta$.  In particular, 
the value of 
$\beta_c$ saturated in this regime must agree with ordinary Yang-Mills theory on $\R^3 \times \S^1_\beta$ 
due to the finite temperature version  of large-$N$ equivalence. 

In Fig.\ref{fig:phase}, we plotted the simplest possible phase diagram of the theory. The semi-classical  analysis is reliable in the $LN\Lambda \lesssim 1$ domain.  We extrapolated the semi-classical result up to  $LN\Lambda/2 \pi = e^{-1} \sim  0.367$ where semi-classical function reaches to its local maximum. \footnote{It is reasonable  to ask up to what value of $LN\Lambda$ one may expect that the semi-classical analysis  will provide an accurate description.  We guess (but do not have a solid  argument)  that this will be the case up to  $LN\Lambda \sim 2 \pi e^{-1} \approx   2.31$.  
 This question is in principle answerable by 
simulating deformed Yang-Mills theory and comparing it with our semi-classical results.}

In the strongly coupled  $LN\Lambda \gg 1$ domain, the transition temperature must be a constant due to large-$N$ volume independence.  Matching the transition temperature to the one of semi-classical analysis at the boundary  of its region of validity, we obtained the phase diagram Fig.\ref{fig:phase}. It should be stated that in this phase diagram, the conjectural region  is  the vicinity of the matching point. Given  $T_c$ at a value few times larger than the matching point,  its $L$-independence at leading order in $N$ is dictated by  volume independence.

We also quote the numerical value for deconfinement temperature at $LN\Lambda/2 \pi = e^{-1}$ which we call the boundary of semi-classical window.  If we use as the strong  scale  
that of QCD, 
$\Lambda_{\rm \overline MS}= 213$MeV, 
this gives us an estimate for $SU(N)$ pure gauge theory 
\begin{equation}
T_c = 0.74 \Lambda \approx  158 
 {\rm MeV}
\end{equation} 
 which is in the same ball-park with the  lattice results, quoted in the Introduction.    \footnote{
 Fixing  the strong scale of Yang-Mills theories with $n_f$ fermions,  one finds, 
 for  $n_f = O(N^0) $, that    $T_c (n_f)  =  T_c^{\rm dYM} /(1 - \frac{2n_f}{11N})$. 
 Numerically, these are      $T_c (n_f=1 ) \approx  168 \;  {\rm MeV} $,  $T_c (n_f=2 ) \approx  180  
 \; {\rm MeV}$,  
  $T_c (n_f=1 ) \approx  193  \; {\rm MeV}  $.  Since for $n_f \geq 1$, the center symmetry is no longer an exact symmetry,  the phase transition is replaced by a rapid-crossover.
  }


A final remark is in order for the $ SU(2) $ theory. Substituting $\alpha_1 = -\alpha_2=1$  in (\ref{eq:Sdual2}) and (\ref{eq:Sdual3}), we observe that the discussion reduces to the one given in Ref.\cite{Kovchegov:2002vi,Dunne:2000vp} for the 3d Georgi-Glashow model up to a trivial rescaling of the fugacity. (For $SU(N)$ with  $N\geq 3$, this is no longer the case, the analog of the $\alpha_N$ monopole, which is on the same footing with $\alpha_1, \ldots, \alpha_{N-1}$ does not exist in a locally 3d theory.) Ref.\cite{Kovchegov:2002vi,Dunne:2000vp} argue that critical point resides in the  $ \beta_m < \beta < \beta_e $ interval. They exhibit, by using  fermionization, that the spectrum has  a massless particle at criticality, and is of Ising universality. This agrees with universality arguments \cite{Svetitsky:1982gs} and the numerical lattice studies  for the  $SU(2)$ pure YM theory on $\R^3 \times \S^1_\beta $, and we expect the deconfinement transition  to remain second order as the radius of $\S^1_L$ is increased. (See Fig.\ref{fig:phase}). For the $SU(N)$ case with $N\geq 3$,  we were not able to determine the order of transition with confidence. We leave this for future work.

\section{Conclusions}
\label{conc}

 We introduced new techniques which enable us to continue 
the deconfinement transition of pure Yang-Mills theory to a  calculable semiclassical domain. This was achieved by exploiting  the recent developments in  large-$N$ volume independence  and semi-classical confinement   in  gauge theories on  $\R^3 \times \S^1_L$  \cite{Shifman:2008ja, Unsal:2008ch, Poppitz:2009uq}. 

Our approach uses a toroidal compactification of gauge theory on $\R^2 \times \S^1_L \times \S^1_\beta$, where at long-distances, the theory reduces to two dimensional field theory. 
A striking feature of this approach is that the deconfinement transition is manifestly seen to be due to a competition between magnetic and electric perturbations in the two-dimensional field theory.  At high temperatures the electric objects dominate, resulting in a deconfined phase. At low temperatures magnetic objects dominate, resulting in a confined phase.  The order parameter distinguishing the two phases is the Polyakov loop, which is  calculable in our framework away from the transition temperature, in {\it both} phases.  

The picture of the deconfinement transition as due to a competition between electric and magnetic objects, is a pleasing one. Confinement due to monopole instantons in 3d and 
 due to  a magnetic Higgs mechanism in 4d has a  long precedent.   
 More recently, Liao and Shuryak  suggested the competition of electric and magnetic objects as a  
 new way to look at the phase diagram of QCD.  In our calculable deformation of Yang-Mills theory, their  scenario is explicitly realized, at least in the semi-classical domain.  

 In the  semi-classical regime, both  electric and magnetic perturbations  are  relevant in the 
 \begin{equation} 
 \frac{T_c}{2} <T < 2T_c, \qquad   {\rm equivalently } \qquad  \beta_m<\beta < \beta_e, 
 \label{regime2}
 \end{equation}
 window. 
In this regime, the densities of both electric and magnetic components of the plasma are comparable, while for $T>2T_c$ the contribution of magnetic objects to the plasma is negligible. Could this window extrapolate to $\R^3 \times \S^1_\beta$ of Yang-Mills theory? Below, we assume this logical possibility and speculate on its consequences. 

In Ref. \cite{Liao:2006ry} a plasma of electric and magnetic charges was studied using a classical molecular dynamics simulation with a variable electric to magnetic density ratio. The measured shear viscosity and diffusion constant were found to be lowest when the densities of electric and magnetic components are equal, and increased otherwise. Comparable densities of electric and magnetic components arise naturally in our field theoretic description in the window (\ref{regime2}). Our analysis  implies that at $T \gtrsim 2T_c$, the magnetic 
perturbations become irrelevant.  In this domain, magnetic excitations are confined to neutral molecules. If the model of Ref.\cite{Liao:2006ry} is a reasonable description of the relevant features of the quark-gluon plasma, and if we extrapolate our semi-classical results to the strong coupling domain, then we expect that for temperatures $T>2T_c$ (which will be probed at  ALICE detector at the Large Hadron Collider at CERN) a rapid increase of shear viscosity and diffusion constant with respect to RHIC results.

\subsection{Open problems}

There are many interesting directions that arise from our construction.  Here, we sort a few which are most pertinent:

\vspace{3mm}

\noindent $\bf 1)$ It would be interesting to study the effect of fermionic   matter  on deconfinement in the semi-classical domain.  In the presence of fermionic matter, the  index theorem on $\R^3 \times S^1_L$ 
\cite{Nye:2000eg, Poppitz:2008hr} implies that  the mechanism of confinement 
is no longer necessarily due to simple monopoles, but rather due to magnetic bions and 
other non-self dual topological defects. (A classification of confinement mechanisms in semi-classical domain is given in \cite{Poppitz:2009uq}.)
\footnote{The index theorem of Refs.\cite{Nye:2000eg, Poppitz:2008hr}
is a variant of
the well-known APS index for  Dirac operator on   $\R^3 \times
\S^1_L$,   a manifold with boundary.  The index of a 4d instanton is a
sum of indices of $N$-types of 3d monopole instantons, ${\cal I}_{\rm
4d, inst.} = \sum_{i=1}^{N}   {\cal I}_{{\cal M}_{\alpha_i}} $, each
of which carry fractional topological charge $1/N$ in the center
symmetric background. A recent lattice simulation
\cite{Hollwieser:2010mj} calculates the index  for some  topological
configurations  and gives evidence for the existence of fractional
topological charge objects.   Outside the semi-classical window, the
index theorem is still valid. However, since the topological defects
are
no longer   dilute at large $LN\Lambda$ or a large four-torus (which
is suitable for lattice studies), it may be harder
to probe fractional topological charge  defects in this domain, see
for example, \cite{Fodor:2009nh}.
Interestingly, the index obtained from lattice
\cite{Hollwieser:2010mj}  and the one in   \cite{Poppitz:2008hr}
agrees.
This is non-trivial and merits  further study. }

\vspace{3mm}

\noindent $\bf 2)$   It would be useful to understand the order of phase transition in the  $NL\Lambda\lesssim1$ regime, and if possible, in the  $LN\Lambda \sim  1$ domain for $SU(N)$ with $N \geq 3$, both numerically and analytically. 

\vspace{3mm}

\noindent $\bf 3)$  It would be interesting to generalize  to orthogonal, symplectic and exceptional  gauge groups, and   in particular,  to the groups for which the cover group has a trivial center symmetry, such as $G_2$.  

\vspace{3mm}

\noindent $\bf 4)$ We have predictions for the $L$ dependence of the 
  critical temperature $\beta_c(L)$  (\ref{critical}), and   mass gap (\ref{massgap})  \cite{Unsal:2008ch}
in the semi-classical domain, and for their $L$-independence in the $LN\Lambda \gg 1$  domain. 
It would be very interesting to test both regimes  in lattice gauge theory and see up to what value of $LN\Lambda$ the semi-classical  description is in good agreement  with lattice results, and at what value of $LN\Lambda$ volume independence sets in. 

\vspace{3mm}

\noindent {\bf 5) } Supersymmetric gauge theories compactified on $\R^3 \times \S^1_L$  are 
 not expected to have any phase transition as a function of $L$\cite{Aharony:1997bx}. 
 A sub-class of supersymmetric theories such as pure $\N=1$ SYM, with gauge boson and adjoint fermion $(A_{\mu}, \lambda_{\alpha})$, \footnote{$\N=1$ mass deformation of $\N=2$ or $\N=4$ SYM will work similarly.}   also possess a semi-classical window in the $LN \Lambda \lesssim 1$ domain where confinement can be shown analytically.  
 It would be useful to understand   how deconfinement sets in when one consider this class of theories on  $\R^2 \times \S^1_L \times \S^1_\beta$, with periodic boundary conditions for bosons and mixed 
\begin{eqnarray} 
&& \lambda(x_1, x_2, x_3 + \beta, x_4) =  - \;  \lambda (x_1, x_2, x_3 , x_4), \cr
&&\lambda  (x_1, x_2, x_3, x_4 + L) = +   \;  \lambda(x_1, x_2, x_3 , x_4).  
\end{eqnarray}
boundary conditions for fermions.  It may also be useful to understand how imposing periodic (supersymmetry preserving) boundary conditions in all  directions avoids  the  phase transition.

\appendix

\section{Yang-Mills in chiral basis and topological defects}
\label{chiral}
In this appendix, we remind the reader the topological defects  pertinent to locally four dimensional gauge theories, in particular to  $\R^3 \times \S^1_L$. It is useful to express the Yang-Mills   action in a chiral basis   which makes the role of self-duality manifest. 
We define $\tau$ and the chiral field strengths $ F_{\pm} $ (which furnishes 
$(3,1) \oplus (1, 3)$ representation of  the Euclidean Lorentz group $SO(4) \sim SU(2)_L \times SU(2)_R$)  as 
\begin{equation}
\tau  = \frac{4 \pi i}{g^2} + \frac{\theta}{2\pi},   \qquad F_{\pm\mu \nu} =  F_{\mu \nu} \pm  \widetilde F_{\mu \nu}  =  F_{\mu \nu} \pm \half \epsilon_{\mu\nu \rho \sigma} F^{\rho \sigma}
 \; .
\end{equation} 
 The Yang-Mills action (\ref{eq:cont}) can be rewritten as 
\begin{equation}
    S^{\rm YM}= \int
     \frac{i}{32 \pi  } \left(  \overline \tau  \;  \tr F_{+\mu\nu}^2 -   \tau \;  \tr F_{-\mu\nu}^2 \right) 
\label{eq:chiral}
\end{equation} 
The $\tau \rightarrow  i \infty $ limit is the weak coupling limit.  In the chiral basis, the instanton equation reads
\begin{equation}
F_{-\mu \nu}=0 \qquad {\rm  or } \qquad   F_{\mu \nu} =   \widetilde F_{\mu \nu} 
\end{equation}
For a 4d instanton, $F_{+\mu\nu}^2= 4F^2_{\mu\nu} = 4 F_{\mu\nu}\widetilde F_{\mu\nu}$, and topological charge  is $ \frac{1}{16 \pi^2} \int \tr  F_{\mu\nu} \widetilde F_{\mu\nu} = 1$. 
 Its  action and $\theta$ angle dependence  appears as 
\begin{equation*}
S=  \frac{i}{32 \pi  } \int \overline \tau  \;  \tr F_{+\mu\nu}^2  = \frac{i }{8 \pi  } 
\int  \overline \tau  \;  \tr  F_{\mu\nu}  \widetilde F_{\mu\nu}  =  2\pi i \overline \tau
\end{equation*}
Thus, in the semi-classical expansion in 4d, the amplitude appears as
\begin{equation}
e^{-S_I} = e^{-2 \pi i \overline \tau } =  e^{- \frac{8\pi^2}{g^2 } - i \theta}
\end{equation}

On small  $\S^1_L \times \R^3$, due to the center-symmetric Wilson line  (\ref{vev}) associated with the boundary  $|\bf x| \rightarrow \infty$, , there are more solutions to $F_{-\mu\nu}+ O(g^2)=0$, 
$F_{ij} - \half \epsilon_{ijk} D_kA_4 + O(g^2)=0, \; i, j, k=1,2,3$,
where $O(g^2)$ part, which is there due to deformation and one-loop potential, is omitted in weak coupling. The  magnetic  and  topological charges of these $N$-types of monopole instantons   are 
given in  (\ref{eq:mag-charge}). In the semi-classical expansion, the amplitudes associated with these instantons are 
\begin{equation}
{\cal M}_{\alpha_i} =  e^{- 2 \pi i \overline \tau/N  } e^{i 
\alpha_i \cdot \bm \sigma}
 =  e^{- \frac{8\pi^2}{g^2N } - i \frac{\theta}{N}} e^{i  \alpha_i \cdot \bm \sigma}
\end{equation}
Notice that, the 4d instanton on   $\R^3 \times S^1_L$ 
may be viewed as a composite of these $N$-types of  3d instantons associated with simple roots $\alpha_i =1, \ldots, N-1$ and the twisted-instanton associated  with affine root $\alpha_N$. These are   sometimes referred to as ``fractional instantons". 
 The  
 corresponding amplitudes obey 
\begin{equation}\prod_{i=1}^{N} {\cal M}_{\alpha_i} =  e^{-2 \pi i  \overline \tau } =
e^{- \frac{8\pi^2}{g^2 } - i \theta }  \, .
\end{equation}
The semi-classical expansion on a center symmetric background is an expansion in 
 $e^{2 \pi i \overline \tau/N  }  =  e^{- \frac{8\pi^2}{g^2N } + i \frac{\theta}{N}}$. The 4d instanton appears in this expansion at $N^{\rm th} $ order.    In particular, at large-$N$, instantons are suppressed  as  $e^{-S_I}= e^{-O(N^1)}$ whereas the fractional instantons of center-symmetric background are $e^{-S_{{\cal M}_{\alpha_i} }}= e^{-O(N^0)}$, hence they are part of large-$N$ dynamics.


\bigskip
\centerline{\bf{Acknowledgements}}

We thank Erich Poppitz, Steve Shenker, Edward  Shuryak, Dam T. Son, Bayram Tekin, Edward Witten and Larry Yaffe for enlightening discussions and comments. M.\"U. thanks Aspen Center for Physics and Weizmann Institute of Science for hospitality, where parts of this work is done. M.\"U. and D.S. are supported by the U.S.\ Department of Energy Grant DE-AC02-76SF00515. D.S. is also supported by the Mayfield Stanford Graduate Fellowship and the Stanford Institute of Theoretical Physics.
\bigskip

\appendix

\end{document}